\documentstyle[emulateapj,onecolfloat]{article}

\input epsf

\def\arcs{\ifmmode {^{\scriptscriptstyle\prime\prime}}
          \else $^{\scriptscriptstyle\prime\prime}$\fi}
\def\arcm{\ifmmode {^{\scriptscriptstyle\prime}}
          \else $^{\scriptscriptstyle\prime}$\fi}
\newdimen\sa  \newdimen\sb
\def\parcs{\sa=.07em \sb=.03em
     \ifmmode $\rlap{.}$^{\scriptscriptstyle\prime\kern -\sb\prime}$\kern -\sa$
     \else \rlap{.}$^{\scriptscriptstyle\prime\kern -\sb\prime}$\kern -\sa\fi}
\def\parcm{\sa=.08em \sb=.03em
     \ifmmode $\rlap{.}\kern\sa$^{\scriptscriptstyle\prime}$\kern-\sb$
     \else \rlap{.}\kern\sa$^{\scriptscriptstyle\prime}$\kern-\sb\fi}
\def\pdeg{\ifmmode $\setbox0=\hbox{$^{\circ}$}\rlap{\hskip.11\wd0 .}$^{\circ}
          \else \setbox0=\hbox{$^{\circ}$}\rlap{\hskip.11\wd0 .}$^{\circ}$\fi}
\def\gtorder{\mathrel{\raise.3ex\hbox{$>$}\mkern-14mu
             \lower0.6ex\hbox{$\sim$}}}
\def\ltorder{\mathrel{\raise.3ex\hbox{$<$}\mkern-14mu
             \lower0.6ex\hbox{$\sim$}}}

\input psfig
\lefthead{}
\righthead{}

\begin{document}

\twocolumn[
\title{ Rapid Calculation of Equatorial Rotation Curves }

\author{C.S. Kochanek}
\affil{Harvard-Smithsonian Center for Astrophysics, Cambridge, MA 02138, 
       ckochanek@cfa.harvard.edu }

\begin{abstract}
We derive a simple, fast one-dimensional integral for the equatorial rotation curve 
of a thin disk with surface density $\Sigma(R)$ modeled as a spheroid with axis ratio 
$q$.  The result is simpler than standard expressions even in the limit of an  
infinitely thin disk ($q\rightarrow 0$).
\end{abstract}

\keywords{ galaxies: kinematics and dynamics }
]

\section{Introduction}

Existing expressions for the calculation of the equatorial rotation curves 
of thin disks have undesirable features.  Derivations using
elliptic integrals are complicated and contain a singularity in the
plane which is usually removed by evaluating the rotation curve slightly
above the plane (e.g. Binney \& Tremaine~\cite{Binney87}).  Derivations
in terms of Hankel (Toomre~\cite{Toomre62}) or Stieltjes (Evans \& de Zeeuw~\cite{Evans92})     
transforms are poorly suited for numerical calculation.  Standard
expressions for finite thickness disks require intrinsically
two-dimensional numerical integrals (e.g. Casertano~\cite{Casertano83},
Cuddeford~\cite{Cuddeford93}) or multi-dimensional tables
(e.g. Dehnen \& Binney~\cite{Dehnen98}).  

In conducting some experiments on the compatibility of central dark matter cusps 
with galaxy rotation curves (see, e.g., Moore~\cite{Moore01}), we needed a faster
means of computing equatorial rotation curves including the effects 
of a finite disk thickness.  By faster we mean a simple, non-singular,
non-oscillatory, one-dimensional numerical integral for the rotation curve 
of a disk with an arbitrary surface density profile $\Sigma(R)$. 
The solution was found by considering the equatorial rotation curves of 
flattened spheroids.  Modeling disks as flat spheroids is little used, although 
the mathematical approach was developed by Brandt~(\cite{Brandt60}),
Brandt \& Belton~(\cite{Brandt62}), Mestel~(\cite{Mestel63}), Lynden-Bell \& Pineault~(\cite{LyndenBell78}),
Lynden-Bell~(\cite{LyndenBell89}) and Cuddeford~(\cite{Cuddeford93}).  These
studies show that the equatorial rotation curve can be expressed as an integral
over the Abel transform of the disk surface density, as also outlined in
Binney \& Tremaine~(\cite{Binney87}).  In this paper we follow the same 
development, but then show that the resulting two-dimensional integral can
be reduced to a one-dimensional integral with convenient properties
for numerical calculation.
We will refer
to our solution as the rotation curve of a spheroidal disk, since it is the
rotation curve of a spheroid ($\rho(R^2+z^2/q^2)$) defined by the surface 
density $\Sigma(R)$ of a disk.

\section{The Equatorial Rotation Curve of a Spheroidal Disk}

We develop our result following the procedures outlined in Binney \& 
Tremaine~(\cite{Binney87}) and Cuddeford~(\cite{Cuddeford93}), although
we focus only on the equatorial rotation curve rather than the global
potential.  For a density distribution $\rho(m^2)$ with $m^2=R^2+z^2/q^2$, 
the equatorial rotation curve is simply
\begin{equation}
    v_c^2(R) = 4\pi G q \int_0^R
        { \rho(m^2) m^2 d m \over \left[ R^2 - (1-q^2) m^2 \right]^{1/2} }
\end{equation}
where the density distribution is the Abel transform of the surface density
\begin{equation}
    \rho(m^2) = - { 1 \over \pi q } \int_{m^2}^\infty { \Sigma(R) \over dR^2 }
                            { d R^2 \over \left( R^2 - m^2 \right)^{1/2} }
\end{equation}
for a sufficiently regular surface density.  For example, the Abel transform of an 
exponential disk, $\Sigma=\Sigma_0 \exp(-R/R_d)$, is 
\begin{equation}
     \rho(m^2) = { \Sigma_0 \over \pi q R_d } K_0(m/R_d)
\end{equation}
where $K_0(x)$ is a modified Bessel function, and the rotation curve is
\begin{equation}
      v_c^2 = 4 G \Sigma_0 R_d \int_0^{R/R_d}
            { K_0(u) u^2 du \over \left[ R^2/R_d^2 - (1-q^2) u^2 \right]^{1/2} }
\end{equation}
Cuddeford~(\cite{Cuddeford93}).  For $q=0$ we obtain the standard analytic result by
Freeman~(\cite{Freeman70}).

Our extension to these results is that we can reduce the calculation to a one-dimensional 
integral by reversing the order of integration to find that
\begin{equation}
   v_c^2(R) = { 4 G R \over 1-q^2 } \int_0^\infty d u { d \Sigma \over du }
      \left[ E(\phi,\alpha) - F (\phi,\alpha) \right]
\end{equation}
where $\alpha^2=(1-q^2)u^2/R^2$ and $\sin \phi = \hbox{min}(1,R/u)$.\footnote{The
special functions
$F(\phi,\alpha)=\int_0^\phi d\theta (1-\alpha^2\sin^2 \theta)^{-1/2}$ and 
$E(\phi,\alpha)=\int_0^\phi d\theta (1-\alpha^2\sin^2 \theta)^{1/2}$ are 
elliptic integrals of the first and second kind respectively.}  The integrand 
has an integrable, logarithmic singularity at $\alpha=1$ ($u=R$) for an infinitely
thin disk ($q\rightarrow 0$).  The result simplifies further to
\begin{equation}
   v_c^2 = -{ 4 G \over 3 R } \int_0^\infty d u { d \Sigma \over du }
      u^2 \sin^3\phi  R_D(\cos^2\phi,1-\alpha^2 \sin^2\phi,1)
\end{equation}
if we use the Carlson forms for elliptic integrals\footnote{Carlson's elliptic integral of the
second kind is defined by
$R_D(x,y,z)={3 \over 2}\int_0^\infty { d\lambda 
    \over (\lambda+x)^{1/2}(\lambda+y)^{1/2}(\lambda+z)^{3/2} }$.}  
rather than the Legendre forms.  The function $R_D(x,y,z)$ has a logarithmic
singularity when $x=y=0$, again corresponding to $\alpha=1$ ($u=R$) with $q=0$.
Integrating by parts, to convert $d\Sigma/du$ into $\Sigma$, and taking the limit
$q\rightarrow 0$ leads to the standard expression for the circular velocity
in terms of elliptic integrals evaluated in the disk plane 
(e.g. Binney \& Tremaine~\cite{Binney87}). Integrands based on $d\Sigma/dR$,
like eqns. (5) and (6), have a weaker singularity than those based on 
$\Sigma(R)$, and the singularity is easily removed by giving the disk a 
non-zero axis ratio $q$.

\section{Summary}

Our expressions for the equatorial rotation curves of spheroidal disks of surface
density $\Sigma(R)$ and axis ratio $q$ are simple (even compared to standard 
expressions for infinitely thin disks) and can be computed very rapidly due to
the excellent numerical properties of the Carlson elliptic integrals.  
The one disadvantage of the result is that real disks (e.g. de Grijs~(\cite{deGrijs98})  
and references therein) are better modeled as separable functions of $R$ and $z$ 
(e.g. $\Sigma(R)\exp(-|z|/H)$) than as spheroids ($\rho(R^2+z^2/q^2)$) with axis
ratio $q$.  This distinction will be important in some circumstances (e.g. 
detailed studies of vertical equilibrium in stellar disks), but should matter
little for the global rotation curve if the axis ratio is adjusted to roughly
match the central disk scale height.  In any case, the simplicity of our final 
expression makes it useful even for infinitely thin disks.  

\acknowledgements Acknowledgments:  I would like to thank J. Binney, W. Dehnen, 
L. Hernquist, H.-W. Rix, and S. Tremaine for considering whether this solution
was previously known.  CSK is supported by the Smithsonian Institution.

\end{document}